%% file: Main.tex
\documentclass[sigconf]{acmart}
\acmConference[ISSTA 2024]{ACM SIGSOFT International Symposium on Software Testing and Analysis}{16-20 September, 2024}{Vienna, Austria}

\AtBeginDocument{%
  \providecommand\BibTeX{{%
    \normalfont B\kern-0.5em{\scshape i\kern-0.25em b}\kern-0.8em\TeX}}}

\usepackage{subfiles} 
\usepackage{multirow}
\usepackage[ruled,linesnumbered]{algorithm2e}
\usepackage{listings}
\usepackage{tcolorbox}
\tcbuselibrary{most}

\usepackage{subcaption}

\usepackage{subfloat}
\usepackage{color,xcolor,colortbl}
\usepackage{xspace}
\usepackage{enumitem}
\usepackage{graphicx}
\usepackage{xr}
\usepackage{url}
\usepackage{booktabs}
\usepackage{diagbox}
\usepackage{array}

\usepackage{pgfplots}
\usepackage{subcaption}

\newcommand{\tool}{SCALE\xspace}

\newcommand{\eg}{\textit{e}.\textit{g}.,\xspace}
\newcommand{\ie}{i.e.,\xspace}

\newcommand{\http}{\url{https://github.com/Xin-Cheng-Wen/Comment4Vul}}

\begin{document}

\title{\tool:
Constructing Structured Natural Language Comment Trees for Software Vulnerability Detection}


\author{Xin-Cheng Wen}
\affiliation{%
  \institution{Harbin Institute of Technology}
  \city{Shenzhen}
  \country{China}}
\email{xiamenwxc@foxmail.com}

\author{Cuiyun Gao$^{\ast}$}
\affiliation{%
  \institution{Harbin Institute of Technology}
  \city{Shenzhen}
  \country{China}}
\email{gaocuiyun@hit.edu.cn}

\author{Shuzheng Gao}
\affiliation{%
  \institution{The Chinese University of Hong Kong}
  \city{Hong Kong}
  \country{China}}
\email{szgao23@cse.cuhk.edu.hk}

\author{Yang Xiao}
\affiliation{%
  \institution{Chinese Academy of Sciences}
  \city{Beijing}
  \country{China}}
\email{xiaoyang@iie.ac.cn}

\author{Michael R. Lyu}
\affiliation{%
  \institution{The Chinese University of Hong Kong}
  \city{Hong Kong}
  \country{China}}
\email{lyu@cse.cuhk.edu.hk}

\thanks{
$^{\ast}$ Corresponding author. The author is also affiliated with Peng Cheng Laboratory and Guangdong Provincial Key Laboratory of Novel Security Intelligence Technologies.
}








\renewcommand{\shortauthors}{Trovato and Tobin, et al.}

\begin{abstract}

\subfile{Sections/0_abstract}

\end{abstract}


\maketitle

\section{Introduction}
\label{sec:introduction}
\input{Sections/1_Introduction}

\section{Background and Related Work}
\label{sec:background}

\input{Sections/2_Background_Motivation}

\section{Proposed Framework}
\label{sec:architecture}
\input{Sections/3_Method}

\section{EXPERIMENTAL Setup}
\label{sec:evaluation}
\input{Sections/4_Evaluation}

\section{Experimental Results}
\label{sec:experimental_result}

\input{Sections/5_Experimental_Result}

\section{Discussion}
\label{sec:discussion}
\input{Sections/6_Discussion}


\section{Conclusion}
\label{sec:conclusion}
\input{Sections/8_Conclusion}

\bibliographystyle{ACM-Reference-Format}
\bibliography{sample-base}










\end{document}

%% file: Sections/0_abstract.tex

Recently, there has been a growing interest in automatic software vulnerability detection.
Pre-trained model-based approaches have demonstrated superior performance than other Deep Learning (DL)-based approaches in detecting vulnerabilities.
However, the existing pre-trained model-based approaches generally employ code sequences as input during prediction, and may ignore vulnerability-related structural information, as reflected in the following two aspects. 
First, they tend to fail to infer the semantics of the code statements with complex logic such as those containing multiple operators and pointers. 
Second, they are hard to comprehend various code execution sequences, which is essential for precise vulnerability detection.
To mitigate the challenges, we propose a {\textbf{S}tructured Natural Language \textbf{C}omment tree-based} vulner\textbf{A}bi\textbf{L}ity d\textbf{E}tection framework based on the pre-trained models, named \textbf{\tool}. The proposed Structured Natural Language Comment Tree (SCT) integrates the semantics of code statements with code execution sequences based on the Abstract Syntax Trees (ASTs).
Specifically, \tool comprises three main modules: 
(1) \textit{Comment Tree Construction}, which aims at enhancing the model's ability to infer the semantics of code statements by first incorporating Large Language Models (LLMs) for comment generation and then adding the comment node to ASTs.
(2) \textit{Structured Natural Language Comment Tree Construction}, which aims at explicitly involving code execution sequence by combining the code syntax templates with the comment tree.
(3) \textit{SCT-Enhanced Representation}, which finally incorporates the constructed SCTs for well capturing vulnerability patterns.
Experimental results demonstrate that \tool outperforms the best-performing baseline, including the pre-trained model and LLMs, with improvements of 2.96\%, 13.47\%, and 3.75\% in terms of F1 score on the FFMPeg+Qemu, Reveal, and SVulD datasets, respectively. Furthermore, \tool can be applied to different pre-trained models, such as CodeBERT and UniXcoder, yielding the F1 score performance enhancements ranging from 1.37\% to 10.87\%. 

%% file: Sections/1_Introduction.tex
Nowadays, modern software development is significantly afflicted by the prevalence of source code security vulnerabilities~\cite{DBLP:journals/ieeesp/McGrawP04}.  
For instance, according to IBM, the year 2023 witnessed a peak in the average cost of data breaches, reaching US\$ 4.45 million~\cite{IBM}.
Over the past decade, there has been a substantial increase in the quantity of identified Common Vulnerabilities and Exposures (CVEs)~\cite{CVE}.  Specifically, during the 2022, the identified CVE number reached 25,227~\cite{Statista}, with a 25.1\% increase over the number of vulnerabilities detected in 2021.  
Therefore, it becomes imperative to develop effective methods for detecting these software vulnerabilities.

Conventional Program Analysis (PA)-based vulnerability detection methods heavily rely on user-defined rules and specifications to identify vulnerabilities, such as INFER~\cite{INFER} and CheckMarx~\cite{CHECKMARX}, which makes them labor-intensive and time-consuming.
Recently, DL-based methods have achieved great success as they can reduce the dependence on expert knowledge and offer a broader spectrum of capabilities in detecting various types of software vulnerabilities~\cite{cheng2022bug}. Early DL-based methods use Convolutional Neural Networks (CNNs)~\cite{wu2017vulnerability, DBLP:conf/icse/WuZD0X022/vulcnn}, Recurrent Neural Networks (RNNs)~\cite{sysevr, russell} and Graph Neural Networks (GNNs)~\cite{devign, reveal} for supervised training. However, the performance of these models can be
largely limited due to the scarcity of vulnerability data~\cite{DBLP:journals/corr/abs-2307-09163/reason}. Recently, the advent of pre-trained models has further advanced this field. These models are trained on massive open-source code repositories and have possessed a vast general programming knowledge. 
The recent studies mainly resort to the pre-training fine-tuning paradigm which injects vulnerability knowledge by further training models on vulnerability datasets. For example, EPVD~\cite{DBLP:journals/tse/ZhangLHXL23/EPVD} and SVulD~\cite{DBLP:journals/corr/abs-2308-11237/svuld} achieve state-of-the-art performance for vulnerability detection.
Although achieving promising performance, the effectiveness of the pre-training techniques is still limited, embodied in two aspects:

\begin{figure}[t]
\centering
 \includegraphics[width=0.47\textwidth]{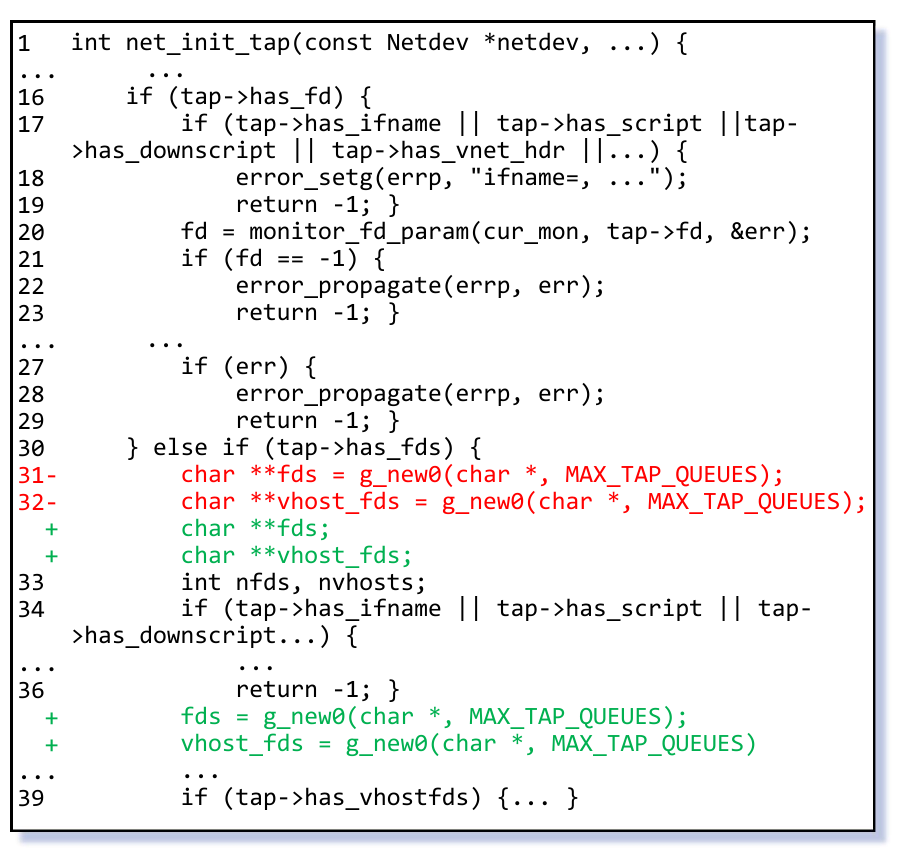}
    \caption{A code example that presents a memory leak vulnerability in the Qemu project, and is misclassified by UniXcoder as non-vulnerable.
    The red and green lines represent the patched code segments before and after fixed, respectively.}
\label{fig:code_example}
\end{figure}

\textit{(1) The pre-trained models tend to fail to infer the semantics of the code statements with complex logic such as those containing multiple operators and pointers.} 
The previous studies~\cite{DBLP:journals/corr/abs-2304-03843/reasongap, DBLP:conf/nips/Wei0SBIXCLZ22/reasongap2} have shown that the ``reasoning gap" in pre-trained models poses a significant challenge. It means that these models struggle to infer the logic of the source code, such as containing multiple operators and pointers, where a single inference error may result in the opposite prediction. Different from natural language, due to the strict rules and syntax, source code may need multiple abstract symbols in a statement to express a simple intent. 
For instance, the code snippet shown in Figure~\ref{fig:code_example} from the Qemu~\cite{Qemu} project potentially causes a memory leak vulnerability in the \texttt{net\_init\_tap()} function. 
In Lines 31-32, the code snippet allocates memory for the arrays of character pointers \texttt{fds} and \texttt{vhost\_fds}. However, 
if any of these error checks failed after the allocation in Lines 34-36, the function would exit without properly deallocating these arrays, leading to a memory leak vulnerability. 
In this example, the complex logic of the source code in Line 34, 
is characterized by the use of multiple operators (\texttt{||} and \texttt{->}), leading to erroneous predictions. It presents challenges for pre-trained models in detecting such vulnerability patterns.
\begin{figure*}
  \begin{subfigure}{0.21\textwidth}
    \includegraphics[width=\textwidth]{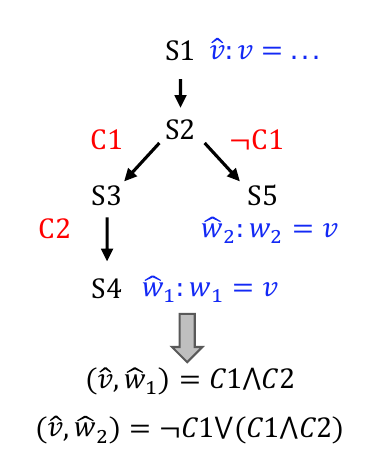}
    \caption{Program analysis methods}
    \label{back::subfigure1}
  \end{subfigure}%
  \hspace{0.5cm}
  \begin{subfigure}{0.18\textwidth}
    \includegraphics[width=\textwidth]{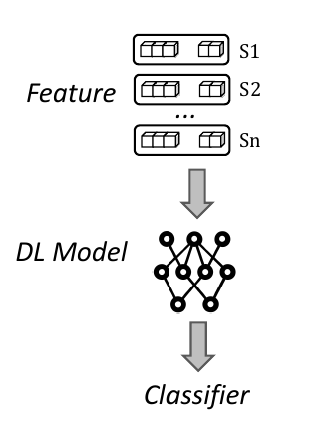}
    \caption{Supervised methods}
    \label{back::subfigure2}
  \end{subfigure}%
  \hspace{0.5cm}
  \begin{subfigure}{0.22\textwidth}
\includegraphics[width=\textwidth]{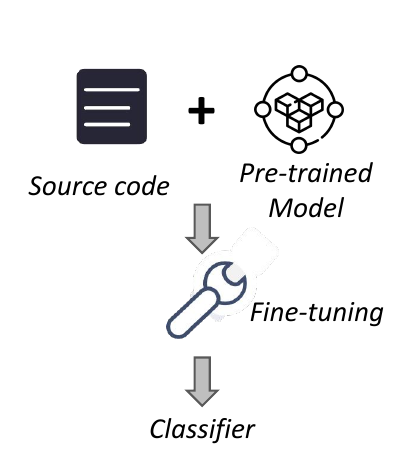}
    \caption{Pre-trained model methods}
    \label{back::subfigure3}
  \end{subfigure}%
  \hspace{0.5cm}
  \begin{subfigure}{0.18\textwidth}
\includegraphics[width=\textwidth]{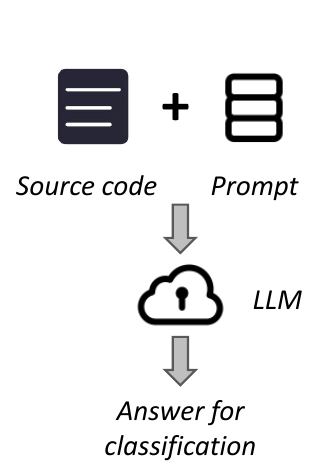}
    \caption{LLM-based methods}
    \label{back::subfigure4}
  \end{subfigure}%
  \caption{Four types of existing vulnerability detection methods.}
  \label{background}
\end{figure*}

\textit{(2) The pre-trained models are hard to capture the code execution sequences.} 
Since pre-trained models primarily utilize code sequences as input, they are hard to comprehend the inherently non-sequential and various code execution sequences of source code, such as commonly-used \texttt{if} and \texttt{return} statements.
For instance, the example shown in Figure~\ref{fig:code_example} contains several \texttt{if} statements and \texttt{return} statements, in which the specific use of \texttt{if} in line 34 and \texttt{return} in line 36 directly causes the vulnerability.
These statements necessitate multiple conditional if statements to trigger a \texttt{return -1}. 
It is challenging for the pre-trained models to 
predict the vulnerability without well capturing code execution sequences.
To address the limitations above, in this paper, we propose a {\textbf{S}tructured natural language \textbf{C}omment tree-based} vulner\textbf{A}bi\textbf{L}ity d\textbf{E}tection framework based on the pre-trained models, named \textbf{\tool}. The proposed Structured Natural Language Comment Tree (SCT) integrates the semantics of code statements with code execution sequences based on the Abstract Syntax Trees (ASTs). \tool has three main modules: 
(1) \textit{Comment Tree Construction}. 
\tool makes the first attempt to incorporate the Large Language Model (LLM)'s general knowledge for comment generation in the vulnerability detection task and further generates the comment tree, which enhances the model's ability to infer the semantics of code statements.
(2) \textit{Structured Natural Language Comment Tree Construction}, which explicitly involves code execution sequence by combining the code syntax templates with the comment tree.
(3) \textit{SCT-Enhanced Representation}, which incorporates the constructed SCTs for well capturing vulnerability patterns.

To evaluate \tool, we use three widely-used datasets in vulnerability detection: FFMPeg+Qemu~\cite{devign}, Reveal~\cite{reveal}, and SVulD~\cite{DBLP:journals/corr/abs-2308-11237/svuld}. We compare \tool with eleven existing vulnerability detection methods. Experimental results demonstrate that \tool outperforms the state-of-the-art baseline, with improvements of 2.96\%, 13.47\%, and 1.17\% in terms of F1 score on the three datasets, respectively. Furthermore, \tool can be applied to different pre-trained models, yielding F1 score performance enhancements ranging from 1.37\% to 10.87\%. These results underscore the effectiveness of \tool in enhancing the detection of software vulnerabilities.

In summary, the major contributions of this paper are summarized as follows:
\begin{enumerate}
\item To the best of our knowledge, we propose a novel structured natural language comment tree, which enhances the model’s ability to infer code semantics by integrating code comments and code syntax templates.
\item We propose \tool, a novel vulnerability detection framework for well capturing vulnerability patterns by enhancing code representation with the structured natural language comment tree.

\item We perform an extensive evaluation 
on three popular datasets, and the results demonstrate the effectiveness of \tool in software vulnerability detection.
\end{enumerate}


%% file: Sections/2_Background_Motivation.tex
We classify the existing vulnerability detection methods into four types: Program analysis-based, Supervised-based, Pre-trained model-based and LLM-based methods, as illustrated in Figure~\ref{background}. 

\subsection{Program Analysis Methods}
Numerous program analysis methodologies have been proposed and demonstrated their effectiveness in vulnerability detection, such as CheckMarx~\cite{CHECKMARX}, FlawFinder~\cite{FLAWFINDER}, PCA~\cite{DBLP:conf/sigsoft/LiCSM20/pca} and RATs~\cite{RATs}. These methodologies typically employ pre-defined rules or patterns to identify improper operations within source code. Figure~\ref{back::subfigure1} illustrates an example of the rules proposed by Saber~\cite{DBLP:conf/issta/SuiYX12/saber}, which use the symbolic rules~\cite{symbolic,DBLP:conf/IEEEares/LiKBL13/symbolic2} to simulate programming logic in source code. However, the creation of well-defined vulnerability rules or patterns heavily relies on expert knowledge~\cite{DBLP:journals/pacmpl/LiTMS18,DBLP:journals/toplas/LiTMS20}, making it challenging to cover all potential cases. Furthermore, the complex programming logic inherent in real-world software projects hinders the manual identification of these rules, undermining the performance of traditional program analysis-based approaches.
\subsection{Supervised Methods}
With the advent of Deep Learning (DL), numerous supervised methods have been proposed that utilize DL-based models to capture program semantics for the identification of potential software vulnerabilities, which mainly include the sequence-based~\cite{DBLP:conf/ndss/LiZXO0WDZ18/vuldeepecker, sysevr} and graph-based~\cite{devign, reveal, IVDETECT} approaches. Figure~\ref{back::subfigure2} illustrates the process of these methods. They typically extract features from source code and adopt the DL models for classification.
For instance, VulDeePecker~\cite{DBLP:conf/ndss/LiZXO0WDZ18/vuldeepecker} and SySeVR~\cite{sysevr} use the bidirectional Long Short-Term Memory (LSTM) network for vulnerability detection and extract the code gadgets from the source code. 

Then, graph-based methods have gained significant popularity due to their interpretability, which has demonstrated more effectiveness than sequence-based methods.
They extract structured representations from the source code, such as AST, CFG, DFG, and Code Property Graphs (CPG)~\cite{DBLP:conf/icait/WangZWXH18/CPGVA}. They then employ Graph Neural Networks (GNNs)~\cite{DBLP:journals/air/WaikhomP23/gnn} models to learn the graph representation for classification. In comparison to traditional program analysis-based approaches, these methods can automatically extract implicit vulnerability patterns from previously vulnerable code, eliminating the need for expert involvement. However, these methods are constrained by semantic embedding and overly complex graph structures.
\subsection{Pre-trained Model Methods}
While GNN-based methods have demonstrated satisfactory performance, they necessitate large and high-quality data for training~\cite{DBLP:journals/corr/abs-2301-05456/dataquality}, which can be challenging to procure in real-world scenarios~\cite{DBLP:journals/corr/abs-2308-10523/PILOT}. An alternative approach involves the use of pre-trained models as the foundational backbone, which are then fine-tuned for the specific downstream task. Figure~\ref{back::subfigure3} illustrates the process of pre-trained model-based methods~\cite{DBLP:conf/iclr/GuoRLFT0ZDSFTDC21/graphcodebert, DBLP:conf/msr/FuT22/linevul,CodeGeeX,DBLP:journals/corr/abs-2401-01060}. These methods, which do not require expert involvement or the generation of structured graphs, utilize source code as input. They then simply fine-tune the pre-trained model and only train the classifier for vulnerability detection. All pre-trained code models with encoder components, such as CodeBERT~\cite{DBLP:conf/emnlp/FengGTDFGS0LJZ20/codebert}, can be utilized to detect vulnerabilities.  CodeT5 ~\cite{DBLP:conf/emnlp/0034WJH21/CodeT5} and UniXcoder~\cite{DBLP:conf/acl/GuoLDW0022/unixcoder} are designed to support both code-related classification and generation tasks. By leveraging the knowledge encapsulated in pre-trained models, these pre-trained model-based approaches achieve the best performance in vulnerability detection. EPVD~\cite{DBLP:journals/tse/ZhangLHXL23/EPVD} proposes an execution path selection algorithm and adopts a pre-trained model to learn the path representations. SVulD~\cite{DBLP:journals/corr/abs-2308-11237/svuld} constructs contrastive paired instances and uses the pre-trained model to learn distinguishing semantic representations.

However, the pre-trained methods generally employ code sequences as input during prediction. They may fail to infer the semantics of the code statement with complex logic and hard to capture the code execution sequence. In this work, we aim to mitigate the issues of pre-trained models for better learning the vulnerability-related structural information.

\subsection{LLM-based Methods}
In recent years, Large Language Models (LLMs) have gained prominence~\cite{DBLP:conf/nips/BrownMRSKDNSSAA20} and widespread adoption in the fields of Natural Language Processing (NLP)~\cite{DBLP:conf/acl-deelio/LiuSZDCC22/nlp} and Software Engineering (SE)~\cite{DBLP:journals/corr/abs-2308-10620/se}. Figure~\ref{back::subfigure4} illustrates the process of these LLMs~\cite{DBLP:journals/corr/abs-2310-09810/llm4vul, ChatGPT}.
Notable examples include the series of GPT models proposed by OpenAI, such as ChatGPT~\cite{ChatGPT} and GPT-4~\cite{GPT4}, as well as the LLaMA models introduced by Meta, including LLaMA~\cite{DBLP:journals/corr/abs-2302-13971/llama} and LLaMA2~\cite{DBLP:journals/corr/abs-2307-09288/llama2}. 

Furthermore, beyond these general-purpose LLMs, specific LLMs trained on code repositories have also achieved parameters in the billion-scale range, such as INCODER~\cite{DBLP:journals/corr/abs-2204-05999/incoder}, StarCoder~\cite{DBLP:journals/corr/abs-2305-06161/starcoder}, and Code Llama~\cite{DBLP:journals/corr/abs-2308-12950/codellama}. 
For example, Code Llama~\cite{DBLP:journals/corr/abs-2308-12950/codellama}, with its substantial 34-billion-parameter model, excels in code generation and completion tasks. 
These models have consistently demonstrated commendable performance across a spectrum of code-related intelligence tasks~\cite{DBLP:conf/kbse/GaoWGWZL23}.

However, these LLMs face significant challenges when applied to software vulnerability detection~\cite{DBLP:journals/corr/abs-2310-09810/llm4vul}. It is mainly caused by individual code snippets often containing numerous undefined identifiers and LLMs often lack domain knowledge for vulnerability detection.

%% file: Sections/3_Method.tex
\begin{figure*}[t]
	\centering
	\includegraphics[width=0.95\textwidth]{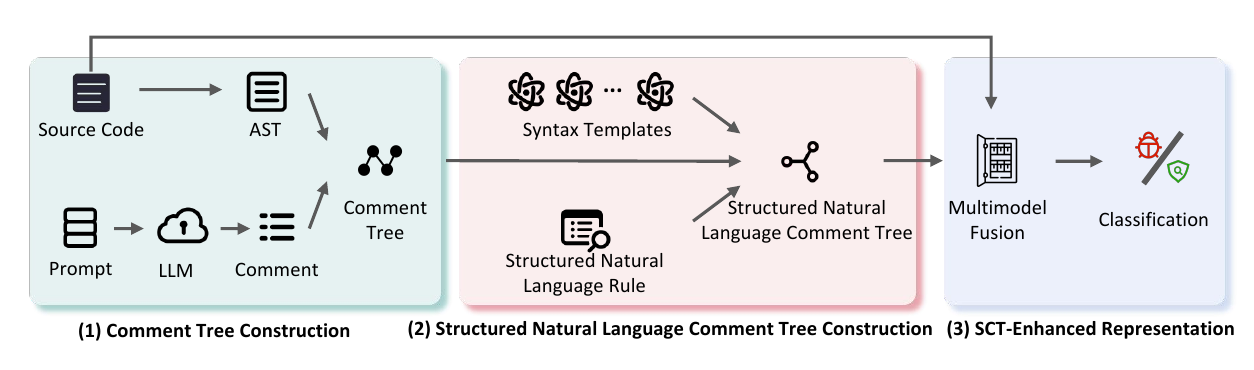}
    \caption{The overview of \tool.}
\label{fig:architecture}
\end{figure*}

In this section,
we elaborate on
the overall architecture of \tool. 
As shown in Figure~\ref{fig:architecture}, \tool mainly consists of three modules: 
comment tree construction,
structured natural language comment tree construction, 
and SCT-enhanced representation, with details as below.

\subsection{Comment Tree Construction}

The purpose of the comment tree construction phase is to enrich corresponding code snippets with natural language descriptions.
It can relieve the difficulties in understanding the semantics of code statements with multiple operators and pointers, and thus help the model infer the code logic more effectively.  
Inspired by the remarkable achievement of LLMs, we seek to utilize their extensive knowledge base to bridge the reasoning gap in vulnerability detection.  Specifically, we leverage ChatGPT to generate comments for each code snippet. Following OpenAI’s gpt-best-practices document~\cite{gpt-best-practices}, we employ the role-based basic prompt to remind ChatGPT~\cite{ChatGPT} of its job (i.e., generate comments ) for each provided code snippet and use the “gpt-3.5-turbo-0301" API for the comment generation.  



To avoid LLMs from generating unnormalized code, 
we then normalize the comments to facilitate the subsequent program analysis part. Specifically, we employ some normalization operations including removing all blank lines
and replacing multi-line comments enclosed in the form of 
triple single quotation marks ($'''$) to 
double slashes ($//$). 
\tool then generates Abstract Syntax Trees (ASTs) based on source code by Tree-sitter~\cite{Tree-sitter} library for constructing the comment tree. A generated AST can be formulated 
as a directed graph $(N, E)$, where $N$ denotes the set of nodes and $E$ represents the set of edges. Each node $n \in N$ within the AST is a representation of a pair $(V, T)$, with $v \in V$ representing the node's value and $t \in T$ representing its type. 
For adding the comment node, we find the first node on the subsequent row and designate this as the target node for the comment. The comment node is then integrated into the AST by taking the parent of the target node as its parent and positioning it as the parent of the target node. 
For example, as shown in Figure~\ref{symbolic_example} and~\ref{fig:sub1}, we generate the comment for Line 3 and traverse the AST of the source code, where the first node in Line 3 is A ``if\_statement''. We add the comment node B as a previous sibling node of A and integrate it into AST.
This approach ensures that the comments are logically and structurally aligned with the following structured natural language rules. 


\subsection{Structured Natural Language Comment Tree Construction}
In this section, we first present our structured natural language rules for structured natural language comment synthesis. Then we describe our proposed algorithm for building SCTs. 
The SCT integrates code comments and code syntax templates, and provides rich code execution information for inferring vulnerability patterns.

\subsubsection{Structured Natural Language Rules} 


\begin{figure}[t]
	\centering
	\includegraphics[width=0.47\textwidth]{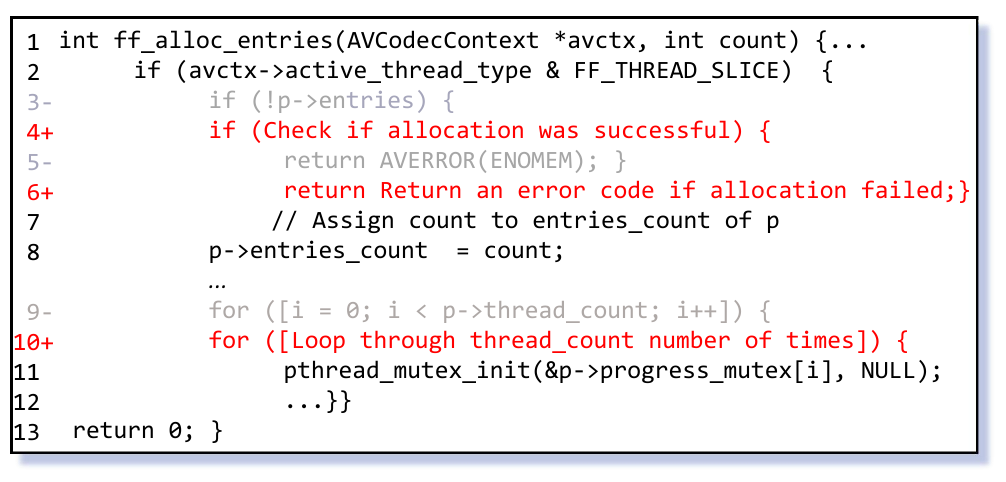}
    \caption{An example of structured natural language comment in \tool. The red and grey lines denote the original source code and structured natural language comment, respectively.}
\label{symbolic_example}
\end{figure}

\begin{figure}
\centering
\begin{subfigure}{.49\textwidth}
  \centering
  \includegraphics[width=\linewidth]{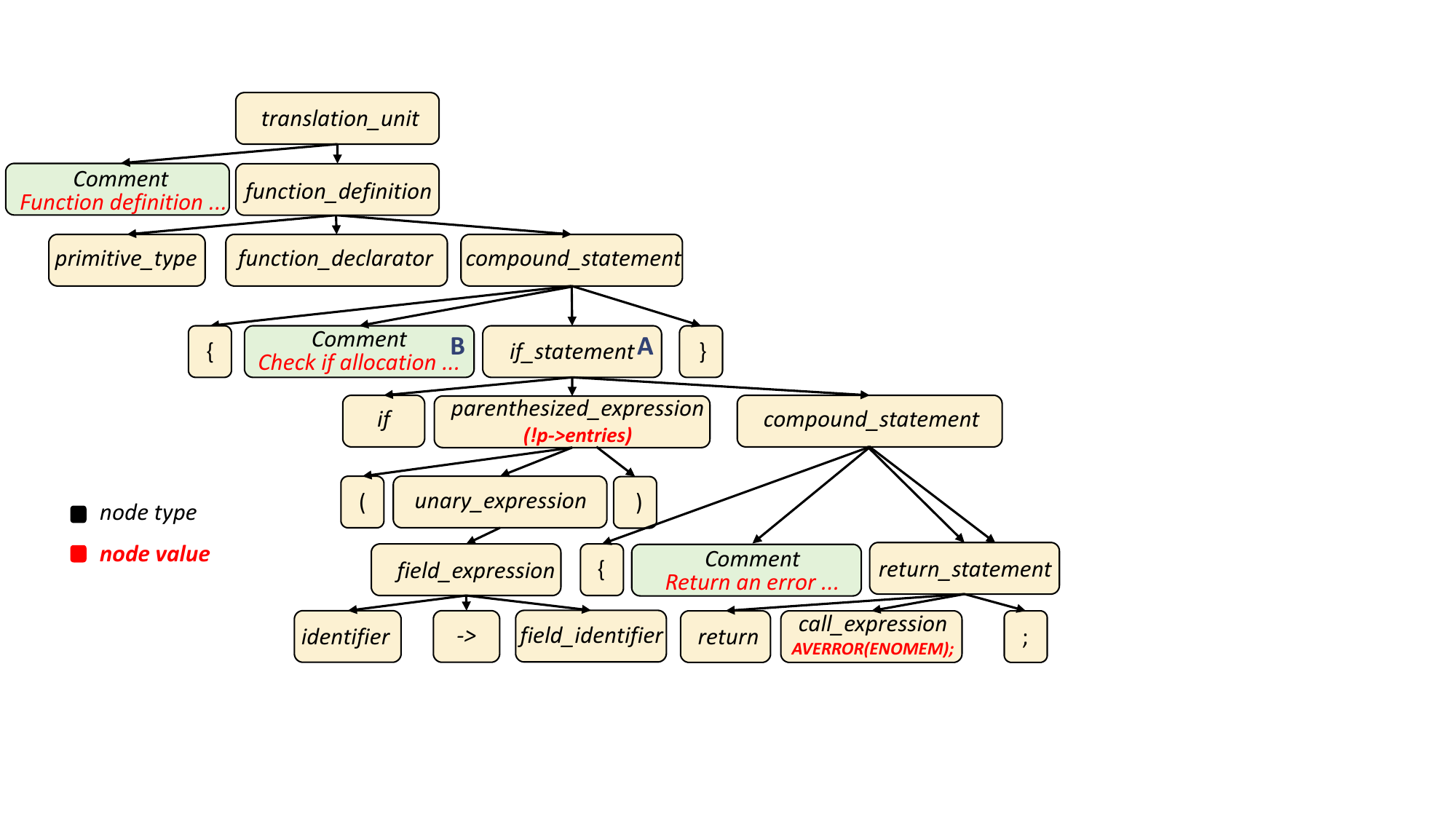}
  \caption{An example of the comment tree.}
  \label{fig:sub1}
\end{subfigure}%
\vspace{0.05cm} 
\begin{subfigure}{.49\textwidth}
  \centering
  \includegraphics[width=\linewidth]{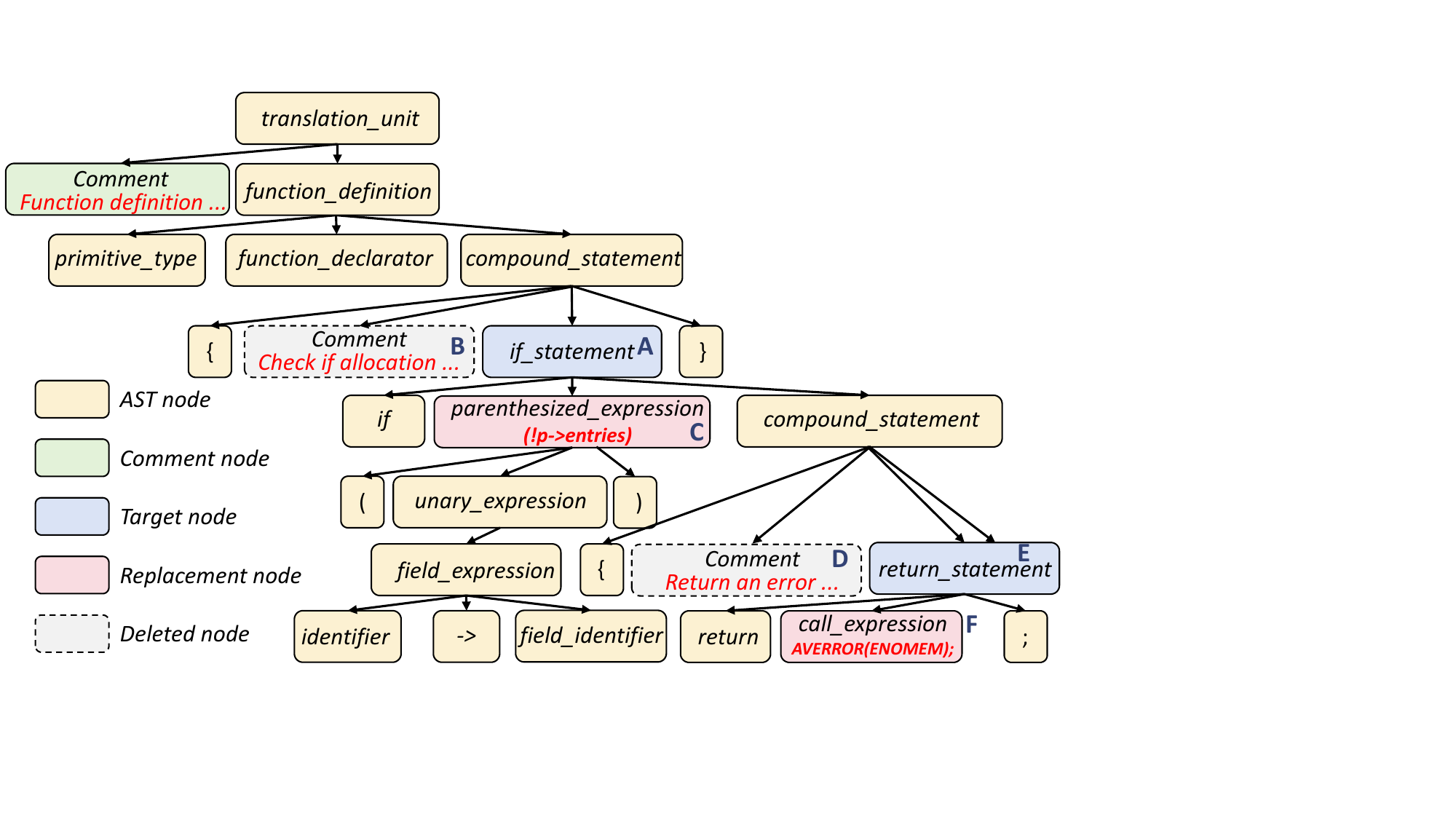}
  \caption{An example of the SCT.}
  \label{fig:sub2}
\end{subfigure}
\caption{The example of the comment tree and SCT. The black and red font denotes the node type and value, respectively. The yellow, green, blue, red, and gray-shaded nodes denote the original nodes, comment nodes, target nodes,
nodes to replace, and deleted nodes, respectively. }
\label{fig:test}
\end{figure}

To integrate
execution sequences with the natural language for more effectively identifying
vulnerability patterns, we introduce a set of structured natural language rules.
Following the guidelines of C++ language reference~\cite{Cppreference}, we select four distinct categories of statements that control how and in what order programs are manipulated and create structured natural language rules for them, including selection statements, iteration statements, jump statements, and labeled statements. The eleven structured natural language rules for these four categories of statements are shown in Table~\ref{rules}. A structured natural language rule represents a type of statement and corresponding target node types.

\input{Tables/Rules}
\textbf{Selection Statements:}
The selection statements encompass ``if'', ``if-else'', and ``switch'' statements, which provide a means to conditionally execute sections of code. 
To construct the structured natural language rules, \tool starts with finding the target node (\eg $if$ and $switch$ ) and corresponding comments description. Then it traverses forward on the target node's child and sibling nodes to confirm the condition branching expression. Finally, these comment descriptions replace the content by utilizing corresponding structured natural language templates.
For the example in Figure~\ref{fig:sub2}, \tool identifies
the target node A $if\_statement$ (highlighted in blue) and extracts the corresponding comment node B (highlighted in grey). It then replaces the child node's C (highlighted in red) value with the comment node B (highlighted in grey).

\textbf{Iteration Statements:}
The iteration statements enable the execution of statements zero or more times, subject to specific termination criteria. We hope to employ natural language rules to facilitate the comprehension of loop logic and triggering conditions by models. 

We design the three structured natural language rules in the second part of Table~\ref{rules}. For the ``while'' statement, \tool starts with the target node type of $while$. Then it traverses forward on the target node's child or sibling nodes to substitute the expression with the corresponding comment. 
The process of constructing structured natural language rules for the ``for'' and the ``range-based-for'' statements closely mirrors that of the ``while'' statement. 
For the ``for'' statement, the replacement encompasses init-expression, condition-expression, and loop-expression, wherein init-expression and loop-expression can contain multiple separated statements.
As shown in Figure~\ref{symbolic_example}, Line 9 (highlighted in grey) indicates the source code of the ``for'' statement, and Line 10 (highlighted in red) indicates the generated structured natural language comments.

\textbf{Jump Statements: }
The jump statements perform an immediate local transfer of control. These statements share common semantics, such as the ``break'' statement, thereby altering the program's execution flow. 
Despite sharing the same tokens, different ``return'' statements possess different implicit targets, often rendering them challenging to discern. Consequently, the structured natural language rules for jump statements can be constructed as intermediaries between the jump statement and the intended inference target, aiding in indicating the code execution sequence
We introduce four categories and the structured natural language templates can be found in the third section of Table~\ref{rules}. 
For the first two structured natural language rules of the jump statements, we find the target nodes of $break$ and $continue$.
Then we combine the initial node value with the corresponding comments to produce
the node value. 
For the $return$ and $goto$ target nodes, we collect the child node value of $identifier$ and $expression$, respectively, where we use the comments to replace the child node values.

In Figure~\ref{fig:sub2}, nodes D to F illustrate an instance of a ``return'' statement. \tool identifies the target node E (highlighted in blue) and substitutes the value of node F (highlighted in red) with that of node D (highlighted in grey). 
This process enriches the ``return'' statement with relevant in-context information, thereby enhancing the model's comprehension of the code
execution sequence.
The generated structured natural language comment is exemplified in lines 5-6 of Figure~\ref{symbolic_example}.

\textbf{Labeled Statements:}
The labeled statements facilitate the direct transfer of program control to a predefined statement. 
To explain the ``case'' statement, we also formulate a structured natural language rule. We capture both the target type node $case$  and its next sibling node from the comment tree. Subsequently, we employ comments to substitute the value of the $constant-expression$'s node. 

\input{Tables/Algorithm}
\subsubsection{SCT Construction Algorithm}
Based on the above structured natural language rules, we then introduce the construction process of SCTs.
The overall SCT construction process is illustrated in Algorithm 1, with an example shown
Figure~\ref{fig:sub2}.

\tool first traverses the comment tree and constructs the SCT using the above rules. 
During this traversal, \tool systematically identifies and records each visited node including statement, expression, and identifier nodes.
When \tool visits a node that conforms to a structured natural language rule, it captures the node's values and incorporates the relevant comment following the structured natural language comments rules. 
According to the parsing rules~\cite{Tree-sitter}, it involves two conditions:
(1) If the target node occurs in the child node, \tool will traverse all child nodes to add structured natural language comments (Lines 4-12). 
(2) If the target node occurs in the sibling node, \tool will check its parent node and only confirm the next sibling node (Lines 13-19). 
Node values meeting these conditions are then updated. This process will be completed until the comment tree is traversed and return the SCT.

Finally, 
\tool flattens the current SCT and generates the structured natural language comment through the AST flatten operator~\cite{Tree-sitter}. Figure~\ref{symbolic_example} shows an example of the flattened SCT
ultimately obtained for model training.

\subsection{SCT-Enhanced Representation}
The SCT-enhanced representation module aims to incorporate the constructed SCTs for well-capturing vulnerability patterns.
We first utilize UniXcoder as the encoder for source code and structured natural language comments, as UniXcoder~\cite{DBLP:conf/acl/GuoLDW0022/unixcoder} is a unified cross-modal (i.e., code, comment, and AST) pre-trained model.
\tool feeds the source code and SCT into the pre-trained model and obtains their representations $h_{c} \in \mathbb{R}^{l \times n}$ and $h_{ct} \in \mathbb{R}^{l \times n}$, where $l$ and $n$ denote the length of input and the dimension of embedding, respectively.
We then propose to fuse the representation of source code and structured natural language comment through Cross-Attention~\cite{DBLP:conf/iccv/ChenFP21/crossattention}. Specifically, \tool first maps the representations into:
\begin{equation}
Q = W_Q^{(i)} \cdot \varphi({h}_{ct}), K = W_K^{(i)} \cdot \varphi({h}_{c}), V = W_V^{(i)} \cdot \varphi({h}_{c})
\end{equation}
where $W_Q^{(i)}$, $W_K^{(i)}$ and $W_V^{(i)}$ are the linear projection of the $i$-th head. 
$\varphi({h}_{ct})$ and $\varphi({h}_{c})$ are the intermediate layer's representation of source code and SCT, respectively. 
\tool integrates the information from source code and SCT, then obtains the SCT-enhanced representation as follows: 
\begin{equation}
Attention(Q,K,V) = softmax\left ( ||_{1}^{H} \frac{QK^{T}}{\sqrt{d}}\right ) \cdot V
\end{equation}
where $T$ denotes the transpose operator and $d$ denotes the embedding size. $||$ and $H$ are the concentrate operator and the head number, respectively.

Finally, \tool uses the mean operator and leverage
a classifier for software vulnerability detection, which 
can be formulated as follows:
\begin{equation}
M = \sigma \left( Classifier\left( Mean (Attention (Q,K,V)) \right) \right)
\end{equation}
where $\sigma$ denotes the sigmoid function. And \tool trains the classifier~\cite{DBLP:journals/tip/HeH0Y16/mlp} by minimizing the Cross-Entropy~\cite{DBLP:journals/anor/BoerKMR05/ce} loss.

%% file: Tables/Rules.tex
\begin{table}[t]
\centering
\setlength{\tabcolsep}{2mm}
\renewcommand{\arraystretch}{1.4}

\caption{
The symbolic rules for constructing the SCT.
``[-]'' indicates the
node value to replace and ``-'' denotes the original AST node type parsed by the Tree-sitter.
}
\resizebox{0.5\textwidth}{!}{
\begin{tabular}{ccl}
\toprule 
\rowcolor[HTML]{DEDEDE}
\textbf{Categories} &\textbf{Target Nodes}             & \multicolumn{1}{c}{\textbf{Symbolic Templates}} \\  
[2pt]
\toprule  
&If            & \textit{\begin{tabular}[c]{@{}l@{}}if ( \textbf{[condition]} ) \\ \qquad if-branch\end{tabular}}                                    \\ 
\cline{2-3}
 &  If-else          & \textit{\begin{tabular}[c]{@{}l@{}}if ( \textbf{[condition]} ) \\         \qquad if-branch \\ else \textbf{[]}\\         \qquad else-branch\end{tabular}} \\
\cline{2-3}
 \multirow{-7}{*}{Selection}&Switch           & \textit{\begin{tabular}[c]{@{}l@{}}switch ( \textbf{[condition]} ) \\ \qquad statement\end{tabular}}                                     \\
\midrule
  &While            & \textit{\begin{tabular}[c]{@{}l@{}}while ( \textbf{[expression]} ) \\ \qquad statement\end{tabular}}                                      \\ 
  \cline{2-3}

   & For              & \textit{\begin{tabular}[c]{@{}l@{}}for (\textbf{[init-expression; condition-expression;} \\ \textbf{ \qquad \qquad loop-expression]}) \\ \qquad statement\end{tabular}} \\ 

  \cline{2-3}
 \multirow{-6}{*}{Iteration}   & Range-based for  & \textit{\begin{tabular}[c]{@{}l@{}}for ( \textbf{[for-range-declaration: expression] }) \\
    \qquad statement\end{tabular}}                \\
\midrule
     &  Break            & \textit{\textbf{[break]};}                                                            \\ 
     \cline{2-3}
&Continue         & \textit{\textbf{[continue]};}                 \\ 
\cline{2-3}
     &Return           & \textit{return \textbf{[expression]} ;}         \\ 
     \cline{2-3}
 \multirow{-4}{*}{Jump}&  Goto        & \textit{goto \textbf{[identifier]} ;}  \\     
 \midrule
 Labeled &Case & \textit{\begin{tabular}[c]{@{}l@{}}case \textbf{[constant-expression]}: \\ \qquad statement\end{tabular}} \\ \bottomrule                                 
\end{tabular}}
\label{rules}
\end{table}

%% file: Tables/Algorithm.tex
\begin{algorithm}[!tpb]
    \SetAlgoLined
    \footnotesize
    \SetKwInOut{Input}{Input}
    \SetKwInOut{Output}{Output}
    \SetKwInOut{Initialize}{Initialize}
    \SetKwFunction{Structured Natural Language Comment Tree}{Structured Natural Language Comment Tree}
    \SetKwProg{Fn}{Function}{:}{}
    \Input{The Comment Tree of given function: $func\_ct$}
    \Output{Structured Natural Language Comment Tree of the given function, $func\_sct$}
    \Initialize{Initialize an action list $ac\_list$ }
    \Fn{Structured Natural Language Comment Tree Construction}{
    

        
    \tcp{Using structured natural language rules to construct structured natural language comment tree} 
    \If{$node$.type $\in$ $ac\_list$ and $node$.comment exist}
    {
        \If{$node$.type $\in$ Table~\ref{rules} and $node$.parent.type == ``comment''}
        {
            \tcp{Processing for child nodes}
            \For{all $child\_node \in node$.children is not visited}
            {
                \If{$child\_node$.type == ``parenthesized\_expression''}
                {
                    st\_comment = $node$.comment \\
                    delete $child\_node$.value \\
                    $child\_node$.value = st\_comment \\
                    update($func\_ct$) \\
                    break \\
                }
            }
            \tcp{Processing for next sibling node}
            \If{next $sibing\_node$ exists and 
                $sibing\_node$.type == ``parenthesized\_expression'' or ``compound\_expression''}
            {
                st\_comment = $node$.comment \\
                    delete $child\_node$.value \\
                    $child\_node$.value = st\_comment \\
                    update($func\_ct$) \\
                    break \\
            }
        }
    }
    $func\_sct = fun\_ct$
    }
    \Return {$func\_sct$}

\caption{Structured Natural Language Comment Tree Construction}
\label{algorithm1}
\end{algorithm}

%% file: Sections/4_Evaluation.tex
\input{Tables/Dataset}

\subsection{Research Questions}

In this section, we evaluate the effectiveness of \tool by comparing it with the state-of-the-art vulnerability detection methods and focus on the following five 
Research Questions (RQs):

\begin{enumerate}[label=\bfseries RQ\arabic*:,leftmargin=.5in]
    \item How effective is \tool in vulnerability detection compared with existing approaches?
    \item How effective is \tool when applied to other pre-trained models? 
     \item How do different
     structured natural language comment rules contribute to the performance of \tool?
    \item What is the influence of different modules on the detection performance of \tool? 
    \item What is the influence of hyper-parameters on the performance of \tool?
\end{enumerate}

\subsection{Datasets}
To answer the questions above, we choose three widely-used vulnerability datasets, including FFMPeg+Qemu~\cite{devign}, Reveal~\cite{reveal}, and SVulD~\cite{DBLP:journals/corr/abs-2308-11237/svuld}. The FFMPeg+Qemu dataset originates from two popular C open-source projects. 
It comprises 22k code snippets, of which 10k have been identified as vulnerable. The Reveal dataset has over 18k code snippets, with around 2k of these exhibiting vulnerabilities. SVulD~\cite{DBLP:journals/corr/abs-2308-11237/svuld} is based on Fan et al.~\cite{DBLP:conf/msr/FanL0N20/fan} and contains both before-fixed and after-fixed code in the training set. The model is required to identify the before-fixed as vulnerable and after-fixed as non-vulnerable simultaneously on this dataset.
Table~\ref{dataset} presents the statistics of the experimental datasets. 

Following the previous work~\cite{DBLP:conf/nips/LuGRHSBCDJTLZSZ21/codexglue, DBLP:journals/corr/abs-2308-10523/PILOT, DBLP:journals/corr/abs-2308-11237/svuld, devign}, we split the datasets into disjoint training, validation, and test sets in a ratio of 8:1:1. We use the training set to train the baseline models, use the validation set for selecting best-performance models, and evaluate the performance of \tool and other baselines in the test set.


\subsection{Baselines}
To verify the effectiveness of \tool, we choose the following three types of 
vulnerability detection approaches as our baselines: 

\begin{itemize}
\item \textbf{Supervised-based methods:} We use Devign~\cite{devign} and Reveal~\cite{reveal}, which are widely adopted as baselines in recent works~\cite{DBLP:journals/infsof/CaoSBWL21, DBLP:journals/corr/abs-2308-10523/PILOT, IVDETECT}. They construct the joint graph from the source code and use the GGNN model for function-level vulnerability detection.

\item \textbf{Pretrained Model-based methods:}
We select three popular pre-trained code models, including CodeBERT~\cite{DBLP:conf/emnlp/FengGTDFGS0LJZ20/codebert}, CodeT5~\cite{DBLP:conf/emnlp/0034WJH21/CodeT5}, and UniXcoder~\cite{DBLP:conf/acl/GuoLDW0022/unixcoder} to detect vulnerabilities. Besides, we also select recent state-of-the-art 
methods that further finetune these pre-trained models for vulnerability detection, including 
EPVD~\cite{DBLP:journals/tse/ZhangLHXL23/EPVD}, LineVul~\cite{DBLP:conf/msr/FuT22/linevul} and SVulD~\cite{DBLP:journals/corr/abs-2308-11237/svuld}.

\item \textbf{LLM-based methods:} We also choose Code Llama-7b~\cite{DBLP:journals/corr/abs-2308-12950/codellama}, ChatGPT~\cite{ChatGPT} and GPT3.5-instruct~\cite{GPT3.5-instruct} to evaluate the performance of LLMs for vulnerability detection. We follow the prompts used in~\cite{DBLP:journals/corr/abs-2310-09810/llm4vul} to improve their performance on vulnerability detection. 
\end{itemize}

\subsection{Evaluation Metrics}
Following the previous work~\cite{devign, DBLP:conf/emnlp/FengGTDFGS0LJZ20/codebert, DBLP:conf/msr/FuT22/linevul}, we choose the following four metrics to evaluate \tool's performance:

\begin{itemize}
\item \textbf{Accuracy (Acc):} $Acc = \frac{TP+TN}{TP+TN+FN+FP}$. Acc measures the percentage of correctly classified samples out of all samples.
$TN$ represents the number of true negatives and $TP+TN+FN+FP$ represents the number of all samples.

\item \textbf{Precision (Pre):} $Pre = \frac{TP}{TP+FP}$. Pre is the percentage of true vulnerabilities out of all the vulnerabilities that are retrieved. $TP$ and $FP$ denote the number of true positives and false positives, respectively.

\item \textbf{Recall (Rec):} $Rec = \frac{TP}{TP+FN}$. Rec is the percentage of vulnerabilities that are detected out of all vulnerable code snippets. $TP$ and $FN$ denote the number of true positives and false negatives, respectively.

\item  \textbf{F1 score (F1):} $F1 = 2 \times \frac{Pre\times  Rec}{Pre+Rec}$. F1 measures the harmonic mean of Pre and Rec metrics.

\end{itemize}

\subsection{Implementation Details}
For all the supervised-based and pre-trained model-based methods except Devign, we directly use the publicly available source code and hyper-parameters released by the authors.   
For Devign, following the previous work~\cite{DBLP:conf/icse/WenCGZZL23, DBLP:journals/corr/abs-2212-14274}, we try our best to reproduce it based on code reproduced by 
other researchers~\cite{reveal}. For LLM-based methods, 
we download the Code-llama-7b from HuggingFace~\cite{HuggingFace} and evaluate them on our server. 
For ChatGPT and GPT-3.5-instruct, we use the OpenAI's public API ``gpt-3.5-\textit{turbo-0301}" and ``gpt-3.5-\textit{turbo-instruct}" for experiments. And we use the initial parameter provided by OPENAI.

To ensure the fairness of the experiments, we use the same data splitting for all the approaches in all research questions. We 
use the ``gpt-3.5-\textit{turbo-0301}" API for the comment generation 
and 
fine-tune the pre-trained model UniXcoder with a learning rate of $2e-5$. The cross attention head number $H$ is set as 8 and the batch size is set to $32$. The influence of these parameters is discussed in Section~\ref{subsec:hyper}. All experiments are conducted on a server with NVIDIA A100-SXM4-40GB GPUs. 

%% file: Tables/Dataset.tex
\begin{table}[t]
\centering
\setlength{\tabcolsep}{2mm}
\renewcommand{\arraystretch}{1.4}

\caption{Statistics of the datasets.}
\resizebox{0.47\textwidth}{!}{
\begin{tabular}{lcccc}

\toprule
\rowcolor[HTML]{DEDEDE}
\textbf{Dataset} & \textbf{\# Total} &  \textbf{\# Vul.}  & \textbf{\# Non-vul.} & \textbf{ Ratio(\%)} \\

\midrule
FFMPeg+Qemu~\cite{devign}  & 22,361   & 10,067 & 12,294   & 45.02  
\\
Reveal~\cite{reveal}  & 18,169   & 1,664  & 16,505   & 9.16          \\
SVulD~\cite{DBLP:journals/corr/abs-2308-11237/svuld} & 28,730  & 5,260 & 23,470  & 18.31        \\
\bottomrule

\end{tabular}
}
\label{dataset}
\end{table}

%% file: Sections/5_Experimental_Result.tex
\input{Tables/RQ1}
\subsection{RQ1: Effectiveness of \tool}
To answer RQ1, we compare the \tool with three types of vulnerability detection baselines.
The results are shown in Table~\ref{RQ1}.

\subsubsection{Comparison with Supervised and Pre-trained Model-based Methods}

Specifically, we compare \tool with two supervised-based 
approaches (i.e., Devign and Reveal) and six pre-trained model-based approaches (i.e, CodeBERT, CodeT5, UniXcoder, EPVD, LineVul, and SVulD). 
From the results in Table~\ref{RQ1}, we observe that \tool 
outperforms all the supervised baseline methods on the three datasets in terms of F1 score, by 2.96\% for FFMPeg+Qemu, 13.47\% for Reveal and 1.17\% for SvulD dataset, respectively. When considering all the four performance metrics in the three datasets (12 combination cases altogether), \tool achieves 
the best performance in 10 out of the 12 cases. 
These results demonstrate that \tool
captures the vulnerability patterns of code more precisely than the supervised and pre-trained model-based baselines.
We also notice that the pre-trained model-based methods perform 
better than supervised-based methods. It can be attributed that pre-trained model-based methods have leveraged general code-related knowledge in the pre-training stage, which enhances the ability to capture vulnerability patterns. 

Moreover, experimental results reveal that the detection accuracy on the SVulD dataset is lower compared to the other datasets. It can be attributed to the challenge that current vulnerability detection approaches struggle to differentiate between before- and after-fixed code which often exhibits only small variances. 
Despite these challenges,  \tool demonstrates improvement in the F1 score of 1.17\%, surpassing the SVulD baseline which is specifically designed for this situation.
\subsubsection{Comparison with LLM-based Methods}
\label{LLMresult}
We also compare \tool with multiple 
LLM-based approaches, including 
Code Llama-7b, ChatGPT and GPT-3.5-instruct. 
In general, we can find 
that the 
LLM-based approaches perform worse than \tool and supervised approaches mentioned before.  
Specifically, \tool 
outperforms the best LLM's baselines by 9.63\%, 170.38\%, 83.83\%, and 150.53\% in terms of accuracy, precision, recall, and F1 score respectively. 
It indicates that despite the extensive model size and training data, LLM-based approaches still face challenges in 
vulnerability detection without 
fine-tuning due to the 
lack of domain knowledge. 

\begin{tcolorbox}
[breakable,width=\linewidth-2pt,boxrule=0pt,top=3pt, bottom=3pt, left=3pt,right=3pt, colback=gray!20,colframe=gray!25]
\textbf{Answer to RQ1:} \tool achieves 
 the best performance in precision and F1 score compared with previous approaches. Specifically, \tool{} outperforms the best-performing baseline by 
 2.96\%, 13.47\%, and 1.17\% in terms of F1 score on the three datasets, respectively. 
 \end{tcolorbox}

\subsection{RQ2: Effectiveness on Other Pre-trained Models}

\input{Tables/RQ2_Pretrain}
In this research question, we wonder whether other pre-trained models can have the same effectiveness when equipped with our proposed SCT.
To answer this question, we replace UniXcoder with different pre-trained model-based methods including CodeBERT and EPVD to validate the performance.
The experimental results are presented in Table~\ref{RQ2}.

The experimental results show that the structured natural language comment is effective for fine-tuning existing pre-trained models in 
most cases.  We find average improvements on different datasets by 6.67\% in CodeBERT, 1.82\% in EPVD and 4.88\% in UniXcoder, 
respectively, in terms of F1 score. In particular, 
incorporating structured natural language comment enables 
CodeBERT to surpass the performance of the majority of existing baselines on the FFMPeg+Qemu dataset. This indicates that SCT provides the extra explanation in structured natural language ways for intricate logic and leverages the structured natural language rules for involving the code execution sequence, which can be easily learned by existing pre-trained model-based methods.
As for EPVD, although it has used multiple syntax control flow paths to capture
vulnerability patterns, 
the structured natural language comment still improves it by alleviating the heavy use of operators and pointers. 
It improves the F1 score performance of 2.69\%, 1.40\%, and 1.37\% in three datasets, respectively.
 
\begin{tcolorbox}
[breakable,width=\linewidth-2pt,boxrule=3pt,top=3pt, bottom=1pt, left=3pt,right=3pt, colback=gray!20,colframe=gray!25]
 \textbf{Answer to RQ2:} 
 SCT is effective for different pre-trained models, which averagely improves the F1 score by 6.67\% in CodeBERT, 1.82\% in EPVD, and 4.88\% in UniXcoder, respectively.
 \end{tcolorbox}
\vspace{3pt}

\subsection{RQ3: Effectiveness of Different Structured Natural Language Rules in \tool}
\input{Tables/RQ3_Rule}
To investigate the effectiveness of each category of rule, we construct four variants of \tool with four categories of structured natural language rules, including selection, iteration, jump, and labeled statements. All variants only use specific structured natural language rules to construct the SCT.

The performance of four variants and \tool is
presented in Table~\ref{RQ3}. The experimental results indicate that the four separate structured natural language rules perform worse
compared with
\tool in most cases. Specifically, we observe that the variants decrease by 0.63\%$\sim$1.06\% in FFMPeg+Qemu and 0.44\%$\sim$0.66\% in Reveal, 1.41\%$\sim$6.21\% in SVulD, respectively, in terms of accuracy.
Compared with other statements, the iteration statements and corresponding structured natural language rules generally contribute more, 
which outperforms other variants in 6 out of 12 cases. We suppose that it is mainly caused by the structured natural language rules of iteration statements replacing more contents (\ie init-expression, condition-expression, and loop-expression) and the constructed structured natural language comments 
can explain the code more precisely.

Besides, we find that leveraging the integration of all structured natural language rules performs better than only using a single rule, which improves the F1 score performance in three datasets by 0.77\%, 1.35\%, and 1.63\%, respectively. This indicates that the integration of all rules can further help the model understand the code's execution sequence to improve the performance of vulnerability detection.

\begin{tcolorbox}
[breakable,width=\linewidth-2pt,boxrule=0pt,top=2pt, bottom=2pt, left=3pt,right=3pt, colback=gray!20,colframe=gray!25]
\textbf{Answer to RQ3:} All 
structured natural language rules contribute to the performance of \tool{}, with an F1 score improvement of 1.41\%$\sim$6.21\%.
Leveraging the integration of all structured natural language rules performs better than using any single rule. 
\end{tcolorbox}

\subsection{RQ4: Effectiveness of Different Modules in \tool}
\input{Tables/RQ4_Ablation}

In this section, we explore the impact of different modules of \tool including Structured Natural Language Comment Tree Construction (SCTC) and SCT-Enhanced Representation (SER) module. The results are shown in Table~\ref{RQ4}.

\subsubsection{Structured Natural Language Comment Tree Construction}
To explore the effect of the SCTC module, 
we deploy one variant (i.e., Without SCTC) by only using the comment tree for SCT-enhanced representation. 
As shown in Table~\ref{RQ4}, the SCTC can improve the performance of \tool on all datasets. 
Specifically, removing structured natural language comments 
leads to the F1 score drop of 1.29\%, 2.59\%, and 2.04\% on different datasets. Especially on the imbalanced dataset (i.e., Reveal), SCTC boosts the performance more than on the balanced datasets (i.e., Devign and SVulD), which enhances the four parametric metrics in Reveal by 0.75\% for accuracy, 2.32\% for precision, 2.87\% for recall, and 2.59\% for F1 score, respectively. It indicates that structured natural language comment generation has a greater effect on the unbalanced dataset.

\subsubsection{SCT-Enhanced Representation}
To understand the impact of the SER module, we also deploy a variant (i.e., Without SER) of \tool without the SER phase. 
In this variant, we only use structured natural language comments as input without the source code.
The performance degradation 
observed across three datasets consistently demonstrate an average decrease 
of 3.42\% in accuracy and 2.83\% in F1 score, respectively.  
These results indicate the SCT-enhanced representation 
can better capture vulnerability patterns. 

\begin{tcolorbox}
[breakable,width=\linewidth-2pt,boxrule=0pt,top=2pt, bottom=2pt, left=3pt,right=3pt, colback=gray!20,colframe=gray!25]
\textbf{Answer to RQ4:} 
Both the SCTC and SER modules enhance the performance of \tool.
The SCG phase improves F1 score performance on three datasets by 1.29\%, 2.59\%, and 2.04\%, respectively.
The SER module improves \tool{} by 1.31\%, 3.61\% and 3.56\%, respectively.

\end{tcolorbox}

\subsection{RQ5: Influences of Hyper-paramaters on \tool}\label{subsec:hyper}
In this section, we study the effect of two hyper-parameters (i.e., batch sizes and head numbers) of \tool. We assign different values to them and examine their impact on the performance of vulnerability detection.
Due to the page limitation, we only present the experimental results of FFMPeg+Qemu in Figure~\ref{RQ5}.. Experimental results for other datasets are shown in the repository.
\subsubsection{Batch Size}
Figure~\ref{RQ5}a shows the performance of \tool across four metrics with different batch sizes in FFMPeg+Qemu. Specifically, the performance of \tool increases with the growth of batch size in general, 
which demonstrates the importance of batch size in improving model generalization.
For batch sizes exceeding 32, we observe that the \tool's performance exhibits relative stability. 
Additionally,  we observe that batch size seriously influences imbalanced datasets (i.e., Reveal), leading to performance fluctuation by 49.43\% $\sim$ 53.60\% in precision and 61.07\% $\sim$ 88.52\% in recall metrics. Conversely, the growth of performance in balanced datasets (i.e., FFMPeg+Qemu and SVulD) is more steady since the similar ratio between negative and positive samples makes the model learning more stable. 
\subsubsection{Head Number}
To evaluate the impacts 
of the head numbers 
of \tool,
we vary it from 2 to 32 and present the performance on four metrics in Figure~\ref{RQ5}b. As can be seen, \tool achieves the best 
accuracy performance when the number of attention heads is set as eight. 
The performance of different head numbers is similar for all datasets. 
Therefore, we empirically use eight head numbers for all three datasets and achieve 65.11\% in FFMPeg+Qemu, 63.59\% in Reveal, and 26.94\% in SVulD, respectively, in terms of the F1 score.

\input{Tables/RQ5_Single}

 \begin{tcolorbox}
 [breakable,width=\linewidth-2pt,boxrule=0pt,top=2pt, bottom=2pt, left=3pt,right=3pt, colback=gray!20,colframe=gray!25]
 \textbf{Answer to RQ5:} 
The hyper-parameter settings can impact the performance of \tool in the FFMPeg+Qemu, Reveal, and SVUlD datasets. \tool achieves the best performance with the batch size of 32 and head number of 8.
 \end{tcolorbox}

%% file: Tables/RQ1.tex
\begin{table*}[h]
\centering
\setlength{\tabcolsep}{2mm}
\renewcommand{\arraystretch}{1.3}
\caption{
Evaluation results of \tool compared with vulnerability detection baselines on the three datasets. 
The shaded cells represent the performance of the best methods in each metric. Bold text cells represent the best performance. The results of statistical significance tests are listed in \http.}
\resizebox{0.92\textwidth}{!}{
\begin{tabular}{l@{\hspace{3\tabcolsep}}cccc@{\hspace{4\tabcolsep}}cccc@{\hspace{4\tabcolsep}}cccc}
\toprule
\rowcolor[HTML]{DEDEDE}
 \multicolumn{1}{l}{\textbf{Dataset}}                     & \multicolumn{4}{c}{\textbf{FFMPeg+Qemu}}                                  & \multicolumn{4}{c}{\textbf{Reveal}}                                                       & \multicolumn{4}{c}{\textbf{SVulD}}                                                        \\
[1pt] 
\rowcolor[HTML]{DEDEDE}
\multicolumn{1}{l}{\textbf{Metrics}}                      & {\textbf{Acc}} & \textbf{Pre} & \textbf{Rec} & \multicolumn{1}{l}{\textbf{F1}}    & {\textbf{Acc}}    & \textbf{Pre}   & \textbf{Rec}      & \multicolumn{1}{l}{\textbf{F1}}          & {\textbf{Acc}}    & \textbf{Pre}   & \textbf{Rec}      & \multicolumn{1}{l}{\textbf{F1}}          \\
[1pt] 
\toprule
Devign                                & 56.89          & 52.50          & 64.67          & 57.95          & 87.49                & 31.55                & 36.65                & 33.91                & 73.57                & 9.72                 & 50.31                & 16.29                \\
 Reveal                                & 61.07          & 55.50          & \textbf{70.70}          & 62.19          & 81.77                & 31.55                & 61.14                & 41.62                & 82.58                & 12.92                & 40.08                & 19.31                \\
\midrule[0.5pt]
CodeBERT                              & 62.37          & 61.55          & 48.21          & 54.07          & 87.51                & 43.63                & 56.15                & 49.10                & 80.56                & 14.33                & 55.32                & 22.76                \\
 CodeT5                                & 63.36          & 58.65          & 68.61          & 63.24          & 89.53                & 51.15                & 54.51                & 52.78                & 78.73                & 14.32                & 62.36                & 23.30                \\
 UnixCoder                             & 65.19          & 59.93          & 59.98          & 59.96          & 88.48                & 47.44                & 68.44                & 56.04                & 77.54                & 15.11                & \textbf{72.24}       & 24.99                \\

 EPVD                                  & 63.03          & 59.32          & 62.15          & 60.70          & 88.87                & 48.60                & 63.93                & 55.22                & 76.75 &  14.26 &  69.58 &  23.67 
 \\
 LineVul     &                          62.37 &	61.55	&48.21&	54.07&	87.51&	43.63	&56.15&	49.10
                & 80.57                & 15.95                & 64.45                & 25.58                \\
 SVulD                                 & -              & -              & -              & -              & -                    & -                    & -                    & -                    & 86.99                & 21.46                & 56.84                & 31.16                \\
\midrule
Codellama-7B                      & 53.44          & 46.20          & 6.37           & 11.20          & 83.49                & 8.23                 & 5.44                 & 6.55                 & 88.40                & 3.84                 & 5.65                 & 4.57                 \\
 ChatGPT                               & 53.85          & 49.72          & 14.35          & 22.28          & 81.77                & 9.52                 & 8.37                 & 8.91                 & \textbf{88.79}       & 14.38                & 25.81                & 18.47                \\
 GPT-3.5-Instruct                   &             51.31 & 47.03 & 42.06 & 44.41 & 60.47 & 9.60 & 32.22 & 14.79 & 65.81 & 7.26 & 50.4 & 12.69 \\
\midrule
\rowcolor[HTML]{f0f0f0}
  \multicolumn{1}{l}{\textbf{\tool}}         & \textbf{66.18} & \textbf{61.88}          & 
68.69          & \multicolumn{1}{l}{\textbf{65.11}} & \textbf{90.02}       & \textbf{52.32}       & \textbf{78.69}                & \multicolumn{1}{l}{\textbf{62.85}}       & 87.63                & \textbf{22.56}       & 57.03                & \textbf{32.33}    \\ 

\bottomrule
\label{RQ1}
\end{tabular}}
\end{table*}

%% file: Tables/RQ2_Pretrain.tex
\begin{table*}[h]
\centering

\setlength{\tabcolsep}{1mm}
\renewcommand{\arraystretch}{1.3}
\caption{The experimental results of applying SCT to
the existing pre-trained model-based methods.
}
\resizebox{1\textwidth}{!}{
\begin{tabular}{lcccc@{\hspace{2\tabcolsep}}cccc@{\hspace{2\tabcolsep}}cccc}
\toprule
\rowcolor[HTML]{DEDEDE}
\textbf{Dataset}                      & \multicolumn{4}{c}{\textbf{FFMPeg+Qemu}}                                  & \multicolumn{4}{c}{\textbf{Reveal}}                                                       & \multicolumn{4}{c}{\textbf{SVulD}}                                                        \\
[1pt] 
\rowcolor[HTML]{DEDEDE}
\textbf{Metrics}                      & \textbf{Acc} & \textbf{Pre} & \textbf{Rec} & \textbf{F1}    & \textbf{Acc}    & \textbf{Pre}   & \textbf{Rec}      & \textbf{F1}          & \textbf{Acc}    & \textbf{Pre}   & \textbf{Rec}      & \textbf{F1}          \\
[2pt] 
\toprule
\textbf{\multirow{2}{*}{CodeBERT\#}}   & 64.24    & 59.07     & 72.11  & 64.94  & 88.78                & 48.30                & 67.62                & 56.41                & 80.06                & 15.28                & 62.74                & 24.58                \\

& ($\uparrow$+1.87\%)    & ($\downarrow$-2.48\%)     & ($\uparrow$+23.90\%) & ($\uparrow$+10.87\%) & ($\uparrow$+1.27\%)                & ($\uparrow$+4.67\%)                & ($\uparrow$+11.47\%)               & ($\uparrow$+7.31\%)                & ($\downarrow$-0.50\%)                & ($\uparrow$+0.95\%)                & ($\uparrow$+7.42\%)                & ($\uparrow$+1.82\%)                \\
[2pt] 
\midrule

\textbf{\multirow{2}{*}{EPVD\#}}       & 61.60    & 56.40     & 72.35  & 63.39  & 89.49                & 63.93                & 50.81                & 56.62                & 76.83                & 74.71                & 15.04                & 25.04                \\
    &($\downarrow$-1.43\%)    & ($\downarrow$-2.92\%)     & ($\uparrow$+10.20\%) & ($\uparrow$+2.69\%)  & ($\uparrow$+0.62\%)    & ($\uparrow$+15.33\%)     & ($\downarrow$-13.12\%) & ($\uparrow$+1.40\%)  &
                            ($\uparrow$+0.08\%)&  ($\uparrow$+60.45\%)& ($\downarrow$-54.54\%)& ($\uparrow$+1.37\%)
 \\[2pt] 
\midrule
\textbf{\multirow{2}{*}{UniXcoder\#}}  & 66.18    & 61.88     & 68.69  & 65.11  & 

89.58	&50.86	&84.84	&63.59
& 82.33                & 17.14                & 62.93                & 26.94                \\
  & ($\uparrow$+0.99\%)    & ($\uparrow$+1.95\%)     & ($\uparrow$+8.71\%)  & ($\uparrow$+5.15\%)  &   ($\uparrow$+1.10\%)                & ($\uparrow$+3.42\%)                & ($\uparrow$+16.40\%)               & ($\uparrow$+7.55 \%)                & ($\uparrow$+4.79\%)                 & ($\uparrow$+2.03\%)                 & ($\downarrow$-9.31\%)                & ($\uparrow$+1.95\%)                 \\
\bottomrule
\end{tabular}
}
\label{RQ2}
\end{table*}

%% file: Tables/RQ3_Rule.tex
\begin{table*}[h]
\centering

\setlength{\tabcolsep}{2mm}
\renewcommand{\arraystretch}{1.3}
\caption{The performance of \tool with different structured natural language rules on three different datasets.
}
\resizebox{0.9\textwidth}{!}{
\begin{tabular}{l@{\hspace{3\tabcolsep}}cccc@{\hspace{4\tabcolsep}}cccc@{\hspace{4\tabcolsep}}cccc}
\toprule
\rowcolor[HTML]{DEDEDE}
\multicolumn{1}{l}{\textbf{Dataset}}                     & \multicolumn{4}{c}{\textbf{FFMPeg+Qemu}}                                  & \multicolumn{4}{c}{\textbf{Reveal}}                                                       & \multicolumn{4}{c}{\textbf{SVulD}}                                                        \\
[1pt] 
\rowcolor[HTML]{DEDEDE}
\multicolumn{1}{l}{\textbf{Metrics}}                      & {\textbf{Acc}} & \textbf{Pre} & \textbf{Rec} & \multicolumn{1}{l}{\textbf{F1}}    & {\textbf{Acc}}    & \textbf{Pre}   & \textbf{Rec}      & \multicolumn{1}{l}{\textbf{F1}}          & {\textbf{Acc}}    & \textbf{Pre}   & \textbf{Rec}      & \multicolumn{1}{l}{\textbf{F1}}          \\
[1pt] 
\toprule
Selection                      & 65.55 & 61.75 & 65.74 & 63.68 & 89.14                         & 49.64                         & 84.02                         & 62.40                         & 80.88                         & 16.22                         & 64.64                         & 25.93                         \\
Iteration                      & 65.12                         & 60.23                         & \textbf{70.84}                         & 65.10                         & 89.18                         & 49.76                         & 84.01                         & 62.50                         & 80.92                         & 16.35                         & 65.21                         & 26.14                         \\
Jump                           & 65.45                         & 61.21                         & 67.65                         & 64.27                         & 88.92 & 49.04 & 83.61 & 61.82 & 80.75                         & 16.22                         & 65.21                         & 25.98                         \\
Labeled                        & 65.37                         & 61.06                         & 67.97                         & 64.33                         & 89.23                         & 49.88                         & 82.79                         & 62.25                         & 76.12 & 13.91 & \textbf{69.58} & 23.19 \\
\hline
\multicolumn{1}{l}{\tool} & \textbf{66.18}                         & \textbf{61.88}                         & 68.69                         & \textbf{65.11}                         & \textbf{89.58}                         & \textbf{50.86}                         & \textbf{84.84}                         & \textbf{63.59}                         & \textbf{82.33}                         & \textbf{17.14}                         & 62.93                         & \textbf{26.94}       \\                 
\bottomrule
\end{tabular}}
\label{RQ3}
\end{table*}

%% file: Tables/RQ4_Ablation.tex
\begin{table*}[h]
\centering

\setlength{\tabcolsep}{1mm}
\renewcommand{\arraystretch}{1.3}
\caption{The performance of \tool in three different datasets when removing the structured natural language comment tree construction module (i.e., Without SCTC) and removing the SCT-enhanced representation module (i.e., Without SER) in four metrics.}
\resizebox{1\textwidth}{!}{
\begin{tabular}{lcccc@{\hspace{2\tabcolsep}}cccc@{\hspace{2\tabcolsep}}cccc}
\toprule
\rowcolor[HTML]{DEDEDE}
\textbf{Dataset}                      & \multicolumn{4}{c}{\textbf{FFMPeg+Qemu}}                                  & \multicolumn{4}{c}{\textbf{Reveal}}                                                       & \multicolumn{4}{c}{\textbf{SVulD}}                                                        \\
[1pt] 
\rowcolor[HTML]{DEDEDE}
\textbf{Metrics}                      & \textbf{Acc} & \textbf{Pre} & \textbf{Rec} & \textbf{F1}    & \textbf{Acc}    & \textbf{Pre}   & \textbf{Rec}      & \textbf{F1}          & \textbf{Acc}    & \textbf{Pre}   & \textbf{Rec}      & \textbf{F1}          \\
[2pt] 
\toprule

\multirow{2}{*}{Without SCTC}       & 65.59    & 61.73     & 66.06  & 63.82 & 89.27    & 50.00     & 75.82  & 60.26 & 78.32    & 15.17     & 69.69  & 24.99 \\
        & ($\downarrow$0.59\%)   & ($\downarrow$0.15\%)     & ($\downarrow$2.63\%)  & ($\downarrow$1.29\%) & ($\downarrow$0.75\%)    & ($\downarrow$2.32\%)     & ($\downarrow$2.87\%) & ($\downarrow$2.59\%) & ($\downarrow$4.01\%)    & ($\downarrow$1.97\%)     & ($\uparrow$6.76\%)  & ($\downarrow$2.04\%) \\
\multirow{2}{*}{Without SER} & 64.28    & 59.68     & 68.53  & 63.80 & 88.96    & 49.02     & 71.72  & 59.24 & 75.02    & 13.90     & 73.57  & 23.38 \\
           & ($\downarrow$1.90\%)    & ($\downarrow$2.20\%)     & ($\downarrow$0.16\%) & ($\downarrow$1.31\%) & ($\downarrow$1.06\%)   & ($\downarrow$3.30\%)     & ($\downarrow$6.97\%)  & ($\downarrow$3.61\%) & ($\downarrow$7.31\%)    & ($\downarrow$3.24\%)     & ($\uparrow$10.64\%) & ($\downarrow$3.56\%) \\
\bottomrule
\multicolumn{1}{l}{\tool}                   & 66.18    & 61.88     & 68.69  & 65.11 & 90.02    & 52.32     & 78.69  & 62.85 & 82.33    & 17.14     & 62.93  & 26.94
\\        
\bottomrule
\end{tabular}}
\label{RQ4}
\end{table*}

%% file: Tables/RQ5_Single.tex
\begin{figure}
    \hspace{-5mm}
  \begin{minipage}{0.25\textwidth}
    \centering
    \begin{tikzpicture}
      \begin{axis}[
        width=\linewidth,
        xtick={1,2,3,4,5},
        xticklabels={4, 8, 16, 32, 64}
      ]
        \addplot coordinates {(1,64.09) (2,64.71) (3,65.52) (4, 66.18) (5, 65.59)};
        \addplot coordinates {(1, 60.94) (2,63.12) (3,61.83) (4, 61.88) (5, 60.78)};
        \addplot [mark=triangle, orange] coordinates {(1,60.80) (2,55.78) (3,65.18) (4, 68.69) (5, 70.76)};
        \addplot coordinates {(1,60.87) (2,59.22) (3,63.46) (4, 65.11) (5, 65.39)};
      \end{axis}
    \end{tikzpicture}
    \label{hyper:1}
    \subcaption{Batch size}
  \end{minipage}
  \hspace{-5mm}
  \begin{minipage}{0.25\textwidth}
    \centering
    \begin{tikzpicture}
      \begin{axis}[
        width=\linewidth,
        xtick={1,2,3,4,5},
        xticklabels={2, 4, 8, 16, 32}
      ]
        \addplot coordinates {(1,65.41) (2,65.56) (3,66.18) (4, 65.01) (5, 65.59)};
        \addplot coordinates {(1, 67.30) (2,60.68) (3,61.88) (4, 59.89) (5, 61.23)};
        \addplot [mark=triangle, orange] coordinates {(1,58.05) (2,71.08) (3,68.69) (4, 72.11) (5, 68.45)};
        \addplot coordinates {(1,56.07) (2,65.47) (3,65.11) (4, 65.44) (5, 64.64)};
      \end{axis}
    \end{tikzpicture}
    \label{hyper:2}
    \subcaption{Head number}
  \end{minipage}
  \caption{The impact of batch size and head number on the model performance in the FFMPeg+Qemu dataset. The blue, red, orange, and black lines denote the accuracy, precision, recall and F1 score metrics, respectively.}
  \label{RQ5}
\end{figure}
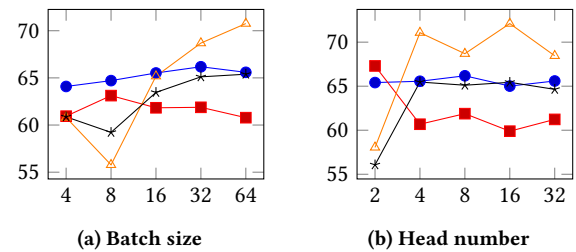

%% file: Sections/6_Discussion.tex
\subsection{Influence of LLM's Generated Comment}
To evaluate the efficacy of the SCT generated by \tool, we further employ GPT-3.5-instruct to construct SCT. Due to constraints in resources, the experiments are limited to the FFMPeg+Qemu dataset in Table~\ref{disscussion}. The results indicate that both GPT3.5-instruct and ChatGPT are capable of producing effective comments for the construction of SCTs. Specifically, the SCTs created by GPT3.5-instruct demonstrated an enhancement in Precision and F1 score, showing improvement of 5.49\% and 1.98\%, respectively, over the best-performing baseline CodeT5, as shown in Table~\ref{RQ1}. In a comparative analysis with SCTs generated by ChatGPT, the GPT3.5-instruct's results are comparably effective. They exhibit superiority in recall and F1 score, whereas ChatGPT outperforms
in accuracy and precision metrics.

These findings suggest that employing different large language models (LLMs) for comment generation on source code 
aids in improving the pre-trained model's comprehension, thereby enhancing the effectiveness of vulnerability detection.

\subsection{Threat to Validity}

\textit{Generalizability on Other Programming Languages.} 
In this paper, we employ the tree-sitter library to parse ASTs for C/C++ programming languages. As a result, our experimental analysis focuses 
solely on C/C++ datasets, excluding other popular languages such as Java and Python. 
In future research, we intend to evaluate the efficacy of \tool in the context of a broader range of programming languages.

\textit{Constraints of Domain Knowledge in Structured Natural Language Rules.}
For the context of C/C++ language reference, \tool offers eleven structured natural language rules to facilitate the correlation between code and its associated comments. Although it is comprehensive enough in most cases, it still can not cover some statements, such as expression statements.
This limitation could potentially introduce biases into the model's predictions. A potential resolution to this issue could involve more 
types of AST nodes to broaden the structured natural language rules, thereby enabling a more thorough integration of comment information.

\textit{Experiments on the larger pre-trained model.}
In this experiment, we evaluate \tool on three pre-trained models. These models are all representative and have shown state-of-the-art performance on benchmarks. However, the size of these models is less than 1B. In the future, we plan to validate the effectiveness of \tool on larger LLMs such as Code Llama~\cite{DBLP:journals/corr/abs-2308-12950/codellama}. 

\input{Tables/Discussion}

%% file: Tables/Discussion.tex
\begin{table}[t]
\centering
\setlength{\tabcolsep}{2mm}
\renewcommand{\arraystretch}{1.1}
\caption{The performance of \tool in FFMPeg+Qemu~\cite{devign} dataset when using
GPT-3.5-Instruct in comment tree construction phase.}
\resizebox{0.47\textwidth}{!}{
\begin{tabular}{lcccc}
\rowcolor[HTML]{DEDEDE}
\toprule
\textbf{Metrics}                      & \textbf{Acc} & \textbf{Pre} & \textbf{Rec} & \textbf{F1}  \\
\midrule
Comment with GPT3.5-Instruct & 63.69                     & 58.23                     & \textbf{74.10}                     & \textbf{65.22}                     \\
\tool (Comment with ChatGPT)        & \textbf{66.18}                     & \textbf{61.88}                     & 68.69                     & 65.11 \\
\bottomrule

\end{tabular}}
\label{disscussion}
\end{table}

%% file: Sections/8_Conclusion.tex
In this paper, we propose \tool, a structured natural language comment tree-based vulnerability detection framework based on the pre-trained models. It mainly comprises a comment tree construction for enhancing the model's ability to infer the semantics of code statements, a structured natural language comment tree construction module for explicitly involving code execution sequence, and an SCT-enhanced representation for well-capturing vulnerability patterns.
Compared with the state-of-the-art methods, the experimental results on three popular datasets validate the effectiveness of \tool for vulnerability detection.
\section*{Data availability}
Our source code and experimental data are available at: \textit{{\http}}.